\documentclass[11pt,a4]{article}
\usepackage{enumitem}
\usepackage{amsthm,amsmath,latexsym,amssymb,amsfonts,amscd}
\usepackage{graphics,lscape,fancyhdr,array,stmaryrd,euscript}
\usepackage{mathabx}
\usepackage{tikz-cd}
\tikzcdset{arrow style=math font}
\usepackage{empheq}
\usepackage{sidecap}
\usepackage{ifpdf}
\usepackage{verbatim}
\usepackage{color}
\usepackage{relsize}
\usepackage{tikz-cd}
\usepackage{titling}
\usepackage{hyperref,setspace}
\usepackage{soul}
\usepackage{cite}
\usepackage{graphicx}
\usepackage{lipsum}
\usepackage{multicol}
\usepackage{yfonts}
\usepackage{stmaryrd}
\usepackage{mathrsfs}

\newcommand{\Ucal}{\mathcal{U}}
\tikzcdset{arrow style=math font}

\definecolor{darkgreen}{RGB}{9, 120, 9}
\definecolor{darkred}{RGB}{120, 9, 9}

\newcommand{\PS}{\mathbb{PS}}

\newcommand{\RR}{\mathbb{R}}

\newcommand{\Bcal}{\mathcal{B}}
\newcommand{\Fcal}{\mathcal{F}}
\newcommand{\Ocal}{\mathcal{O}}
\newcommand{\Ncal}{\mathcal{N}}
\newcommand{\PT}{\mathbb{PT}}

\newcommand{\PPb}{\overline{\mathbb{P}}}
\newcommand{\CC}{\mathbb{C}}

\newcommand{\projectivespace}{\mathbb{CP}}

\newcommand{\MM}{\mathbb{M}}

\newcommand{\Scal}{\mathcal{S}}

\newcommand{\Acal}{\mathcal{A}}

\newcommand{\pl}{\partial}
\newcommand{\pvec}{\boldsymbol{p}}

\newcommand{\Tr}{\text{Tr}}

\begin{document}

\hfill
\vskip 0.01\textheight
\begin{center}
{\Large\bfseries 
Twistor Constructions for Higher-spin Extensions of\\ (Self-dual) Yang-Mills}\\

\vskip 0.03\textheight

Tung Tran\,\footnote{Email: vuongtung.tran@umons.ac.be}

\vskip 0.04\textheight

{Service de Physique de l’Univers, Champs et Gravitation, Université de Mons,\\
20 place du Parc, 7000 Mons, Belgium}

\end{center}

\vskip 0.02\textheight

\begin{abstract} 
We present the inverse Penrose transform (the map from spacetime to twistor space) for self-dual Yang-Mills (SDYM) and its higher-spin extensions on a flat background. The twistor action for the higher-spin extension of SDYM (HS-SDYM) is of $\Bcal \Fcal$-type. By considering a deformation away from the self-dual sector of HS-SDYM, we discover a new action that describes a higher-spin extension of Yang-Mills theory (HS-YM). The twistor action for HS-YM is a straightforward generalization of the Yang-Mills one.
\end{abstract}
\newpage
\tableofcontents
\newpage
\section*{Introduction}

It is well known that many interesting integrable systems, such as self-dual Yang-Mills (SDYM) \cite{Chalmers:1996rq} or self-dual gravity (SDGRA) \cite{Siegel:1992wd}, can be described as (anti)-holomorphic structures on twistor space. One of the advantages of formulating physics in terms of twistor variables is that we can sometimes have a more natural and geometrical interpretation than studying the same physics on spacetime. For instance, supersymmetric SDYM and SDGRA can be formulated as Chern-Simons theories on super twistor space \cite{Witten:2003nn,Boels:2006ir,Mason:2007ct}. Since Chern-Simons theories are topological, SDYM and SDGRA must be quasi-topological (almost free) in spacetime. Indeed, the only non-vanishing amplitudes for both SDYM and SDGRA are the three-point amplitudes (with complex momenta) and the one-loop amplitudes of all-plus helicity \cite{Siegel:1992wd,Bern:1998sv}. Conventional theories such as (super) Yang-Mills theory can also be constructed from twistor space \cite{Mason:2005zm,Boels:2006ir}. In addition, the long-sought twistor action for general relativity formulated in terms of Plebanski action has been discovered recently \cite{Sharma:2021pkl}.

The connection between twistor space and spacetime is known as twistor correspondence. Let us demonstrate it by the following double fibration:
\begin{equation*}
    \begin{tikzcd}[column sep=small]
& \mathbb{PS} \ar["\pi_1" ',dl] \ar[dr,"\pi_2"] & \\
\PT  & & \MM^4
\end{tikzcd}    
\end{equation*}
Here, $\PT$ is the projective twistor space, and $\MM^4$ is the four-dimensional flat spacetime. The space $\PS$ is the corresponding space between $\PT$ and $\MM^4$. To get spacetime dual of a given system in $\PT$, one first uses the pullback $\pi_1^{-1}$ to map data on $\PT$ to $\PS$ and then performs the projection $\pi_2$ (which in most cases is trivial). The map from twistor space to spacetime is known as the Penrose transform. Although it is a well-established operation, it does not always guarantee a covariant expression in spacetime even if we start with a covariant action on twistor space, see \cite{Mason:2007ct} for the case of SDGRA.

In this paper, we consider the inverse map, i.e. the map that lifts a system from spacetime to twistor space (through $\PS$), which we refer to as the inverse Penrose transform. One of the advantages of this approach is that we can start with a covariant action (if available) that has all the desired properties of a field theory in spacetime. Hence, the outcome from the opposite direction, i.e. from twistor space to spacetime, is guaranteed to be covariant. To study the inverse map, we first revisit the cases of self-dual Yang-Mills \cite{Chalmers:1996rq} and Yang-Mills \cite{Boels:2006ir}. Then, we study the higher-spin extension of (self-dual) Yang-Mills \cite{Krasnov:2021nsq}. It is important to note that while the covariant action for SDYM has been known for a long time, the covariant action for SDGRA \cite{Krasnov:2016emc,Krasnov:2021cva} and the higher-spin extensions of SDYM and SDGRA with the correct degrees of freedom have only been obtained recently, see \cite{Krasnov:2021nsq}. It resolves a general belief that one can only consistently write down higher spin theories in flat spacetime in the light-cone gauge. A common feature of these theories is that they have only cubic interactions in flat spacetime with two fields of the same-sign helicities, and another with opposite helicity plays the role of a Lagrangian multiplier.

The higher-spin extension of SDYM (HS-SDYM) considered in this paper is  also known as one-derivative chiral higher spin theory \cite{Ponomarev:2017nrr} that is a closed subset of chiral higher spin gravity (HSGRA) \cite{Metsaev:1991mt,Metsaev:1991nb,Ponomarev:2016lrm,Skvortsov:2018jea}. The chiral HSGRA theory is UV-finite at one-loop \cite{Skvortsov:2020wtf,Skvortsov:2020gpn}. And, we expect it to be one-loop exact. Due to various No-go theorems in flat \cite{Coleman:1967ad,Weinberg:1964ew} and AdS space \cite{Maldacena:2011jn}, see \cite{Bekaert:2010hw} for a review, it has been challenging to come up with viable interacting theories of massless higher spin. In one way or another, the (holographic) S-matrix turns out to be trivial/simple, which indicates that:
\begin{enumerate}
    \item Higher spin symmetry (an infinite-dimensional symmetry) can constrain interactions such that they cancel each other out in the physical amplitudes.
    \item Higher spin theories with local interactions must be either topological \cite{Blencowe:1988gj,Bergshoeff:1989ns,Campoleoni:2010zq,Henneaux:2010xg,Pope:1989vj,Fradkin:1989xt,Grigoriev:2019xmp,Grigoriev:2020lzu} (with no propagating degrees of freedom) or quasi-topological, meaning they have propagating degrees of freedom and non-trivial vertices but appear almost free; or higher-spin extensions of Weyl gravity \cite{Segal:2002gd,Tseytlin:2002gz,Bekaert:2010ky}. 
\end{enumerate}
We showed that the action of HS-SDYM has a similar Chalmers-Siegel form in the light-cone gauge \cite{Krasnov:2021nsq}. It is not surprising since there is a close relationship between the chiral HSGRA and SDYM \cite{Ponomarev:2016lrm,Skvortsov:2018jea,Skvortsov:2020wtf,Skvortsov:2020gpn}. Moreover, we show that the twistor action counterpart of HS-SDYM is a generalization of $\Bcal\Fcal$-type action. It strongly indicates that HS-SDYM must be a quasi-topological field theory. We expect that the chiral HSGRA should also be quasi-topological. One of the purposes of this paper is to construct a twistor action for HS-SDYM (the 'simplest' interacting higher spin theory) and then try to understand whether twistor theory can provide a natural way to obtain a covariant action for the chiral HSGRA.\footnote{The construction of conformal HSGRA using twistorial techniques can be found in \cite{Haehnel:2016mlb,Adamo:2016ple}.} By adding a small perturbation to the HS-SDYM action, we acquire a new theory that describes the higher-spin extension of Yang-Mills, and we call it HS-YM for short. Therefore, one may expect to discover another class of HSGRA by deforming the chiral theories similarly to what we do to SDYM and SDGRA. Hence, the search for a covariant description of the chiral HSGRA (and its truncations) becomes very attractive as we may finally understand if we can have a higher spin theory with \textit{local} higher-point vertices. Yet, if the answer is negative, it will teach us more about the No-go theorems in flat space.

The paper is organized as follows. In section \ref{sec:1}, we first review twistor geometry and the Penrose transform in preparation for the later sections. Moving to section \ref{sec:2}, we warm up by scrutinizing the inverse Penrose transform for SDYM. Then, we study the twistor construction of the full Yang-Mills. Although most of the material in section \ref{sec:2} is standard in twistor theory, the inverse map from spacetime to twistor space for interacting theories is what we want to explore with the hope that it can bring new insight when constructing twistor actions of more complicated cases. In section \ref{sec:3}, we show how one can construct the twistor actions for free scalar and higher-spin fields. These fields are indispensable in the spectrum of generic higher spin theories. Section \ref{sec:4} is devoted to the twistor constructions of HS-SDYM and HS-YM. We conclude with a discussion in section \ref{conclusion}. 
\section{Preliminaries}\label{sec:1}
The twistor space $\mathbb{T}$ is defined to be an open subset of $\CC^4$ with coordinates $Z^A=(\lambda_{\alpha},\mu^{\alpha'}$), where $\mu$ and $\lambda$ are spinors that carry opposite chirality. Here, the Greek indices $\alpha,\beta=0,1$ and similarly $\alpha',\beta'=0',1'$. In general, one prefers to work with the projective version of the twistor space (a subspace of the complex projective space $\projectivespace^3$), denoted $\PT$, in which $Z^A\neq 0$. Then, $\PT=\projectivespace^3\backslash \projectivespace^1$ for $\lambda_{\alpha}\neq 0$.\footnote{In other words, we remove the projective line where $\mu^{\alpha'}\neq0$ and $\lambda_{\alpha}=0$ from $\projectivespace^3$.} On the other hand, our four dimensional spacetime is chosen to be real Minkowski space with Euclidean signature, i.e. $\MM^4=\RR^4$. The twistor correspondence, which relates points in $\PT$ with lines in $\MM^4$ (or inversely lines in $\PT$ with points in $\MM^4$), reads
\begin{align}\label{eq:incident}
    \mu^{\alpha'}=x^{\alpha\alpha'}\lambda_{\alpha}\,.
\end{align}
This algebraic relation is known as the incidence relation and can be presented in terms of a double fibration of the projective spinor bundle over $\MM^4$ and $\PT$:
\begin{equation}\label{doublefibration}
    \begin{tikzcd}[column sep=small]
& \mathbb{PS} \ar["\pi_1" ',dl] \ar[dr,"\pi_2"] & \\
\PT  & & \MM^4
\end{tikzcd}    
\end{equation}
The coordinates on $\mathbb{PS}$ are $(\lambda_{\alpha},x^{\alpha\beta'})$ where $\lambda_{\alpha}\in \projectivespace^1$ (the Riemann sphere). The map $\pi_1:\mathbb{PS}\rightarrow\PT$ is the projection $(\lambda_{\alpha},x^{\beta\beta'})\mapsto (\lambda_{\alpha},\mu^{\beta'}=x^{\beta\beta'}\lambda_{\beta})$, where the incident relation is imposed, while $\pi_2:\mathbb{PS}\rightarrow \MM^4$ is the trivial projection $(\lambda_{\beta},x^{\alpha\alpha'})\mapsto x^{\alpha\alpha'}$. In Euclidean signature, we have the following conjugation of Weyl spinors \cite{Woodhouse:1985id}
\begin{align}\label{eq:hatoperator}
    \lambda_{\alpha}=(\lambda_0,\lambda_1)\mapsto \hat{\lambda}_{\alpha}=(-\overline{\lambda_1},\overline{\lambda_0})\,,\qquad \mu^{\alpha'}=(\mu^{0'},\mu^{1'})\mapsto \hat{\mu}^{\alpha'}=(-\overline{\mu^{1'}},\overline{\mu^{0'}})
\end{align}
Notice that $\hat{\hat{\lambda}}=-\lambda$ (similarly for $\mu$). Now we can introduce a dual twistor $\hat{Z}$ of $Z$ such that $\hat{Z}^A=(\hat{\lambda}_{\alpha},\hat{\mu}^{\alpha'})$. Having a pair of twistors $Z$ and $\hat{Z}$, the inverse of \eqref{eq:incident} reads (see Appendix \ref{app:convention} for our conventions)
\begin{align}\label{eq:inversebosonic}
    x^{\alpha\alpha'}=\frac{\mu^{\alpha'}\hat{\lambda}^{\alpha}-\hat{\mu}^{\alpha'}\lambda^{\alpha}}{\langle \hat{\lambda}\lambda\rangle}\,.
\end{align}
Therefore, we can identify the twistor space $\PT$ with $\mathbb{PS}\cong \MM^4\times \projectivespace^1 \sim \MM^4\times S^2$ since each point in $\PT$ corresponds to a unique point in $\MM^4$ through the incident relation \eqref{eq:incident}. The complex structure on $\PT$ is given by the (anti-holomorphic) Dolbeault operator
\begin{align}
    \bar{\pl}:=d\hat{Z}^A\frac{\pl}{\pl \hat{Z}^A}=d\hat{\mu}^{\alpha'}\frac{\pl}{\pl \hat{\mu}^{\alpha'}}+d\hat{\lambda}_{\alpha}\frac{\pl}{\pl \hat{\lambda}_{\alpha}}\,,
\end{align}
such that $\bar{\pl}\ :\ \Omega^{p,q}(\PT)\mapsto \Omega^{p,q+1}(\PT)$. Clearly, $\bar{\pl}^2=0$. In addition, we also have the (holomorphic) Woodhouse operator
\begin{align}
    \hat{\pl}=d\hat{Z}^A\frac{\pl}{\pl Z^A}\,,
\end{align}
that obeys $\hat{\pl}^2=0$ and $\bar{\pl}\hat{\pl}+
\hat{\pl}\bar{\pl}=0$. More details can be found in \cite{Boels:2006ir,Wolf:2010av,Adamo:2017qyl,Adamo:2013cra}. 

To perform the inverse Penrose transform, one can do the following.\footnote{The algorithm is the same for supersymmetric cases.} We start with a spacetime action on $\MM^4$ and map it to the corresponding space $\mathbb{PS}$. In most cases, this is a trivial insertion of an integral over the fibre $\projectivespace^1$. Then, upon imposing the incident relations, i.e. the $\pi_1$  projection, we obtain a twistor action counterpart of the action in $\MM^4$. The inverse Penrose transform was previously applied in \cite{Eastwood:1981jy,Woodhouse:1985id} at the level of equations of motion. We will, however, work with the inverse Penrose transform at the level of the action. Meaning, we will try to import spacetime actions into twistor space. Note that since $\PS\cong \PT$ in Euclidean signature, it gives us more flexibility to work with the inverse Penrose transform. 

\paragraph{The Penrose transform.} To understanding the inverse Penrose transform, it is helpful to review the concept of the Penrose transform first. One of the goals of twistor theory is to establish the correspondence between field equations in $\MM^4$ and holomorphic structures on $\PT$. For massless fields of any spin in 4d Minkowski, we can represent them as cohomology classes on twistor space via the Penrose transform \cite{Eastwood:1981jy,Atiyah:1979iu,Woodhouse:1985id}. In the case of the spin-1 field, the field strength reads
\begin{align}
    F_{\mu\nu}\mapsto F_{\alpha\alpha'\beta\beta'}=\epsilon_{\alpha\beta}F_{\alpha'\beta'}+\epsilon_{\alpha'\beta'}F_{\alpha\beta}
\end{align}
where $F_{\alpha\beta}$ and $F_{\alpha'\beta'}$ are referred to as the self-dual and anti-self-dual parts of the field strength. The Maxwell equations are then reduced to a single (conformally invariant) equation
\begin{align}\label{eq:spin-1 eq}
    \pl_{\alpha\alpha'}F^{\alpha\beta}=0\,
\end{align}
under this decomposition of the field strength. The above equation \eqref{eq:spin-1 eq} describes massless spin-1 field with, say, negative helicity. The opposite helicity is represented by a gauge potential $A^{\alpha,\alpha'}$ \cite{Eastwood:1981jy,Atiyah:1979iu,Woodhouse:1985id} that obeys
\begin{align}
    \pl^{\alpha}_{\  \alpha'}A^{\alpha,\alpha'}=0\,.
\end{align}
The above equation is invariant under the gauge transformation $\delta A^{\alpha,\alpha'}=\pl^{\alpha\alpha'}\xi$. Here, we use the convention that indices that are symmetrized are denoted by the same Greek letters, e.g. the term $\lambda_{\alpha}\xi_{\alpha}$ denotes $\frac{1}{2}\big(\lambda_{\alpha_1}\xi_{\alpha_2}+\lambda_{\alpha_2}\xi_{\alpha_1}\big)$. We will also denote a fully symmetric rank-$m$ tensor by $T_{\alpha(m)}=T_{\alpha_1...\alpha_m}$. The spacetime action for free spin-1 field takes a simple form
\begin{align}\label{eq:actionfreespin1}
    S=\int d^4x\,\psi_{\alpha\alpha}\pl^{\alpha}_{\ \alpha'}A^{\alpha,\alpha'}\,,
\end{align}
where we have changed $F_{\alpha\alpha}$ to $\psi_{\alpha\alpha}$ to differ from the notation of the usual field strength. Now, let us turn our focus to twistor space and consider the following action
\begin{align}\label{eq:actionfreespin1ontwistor}
    \Scal=\int_{\PT} D^3Z \,\Bcal \wedge  \bar{\pl}\Acal\,,
\end{align}
where $D^3Z$ is the canonical holomorphic measure on $\projectivespace^3$ of weight 4 in $\lambda$ \cite{Witten:2003nn}:
\begin{align}\label{eq:measurePT}
    D^3Z=\epsilon_{ACBC}Z^AdZ^B\wedge dZ^C\wedge dZ^D= \langle \lambda d\lambda\rangle \wedge [d\mu\wedge d\mu]\,.
\end{align}
Here, $\Acal$ is a $(0,1)$-form connection and $\Bcal$ ,which is also a $(0,1)$-form, plays the role of a Lagrange multiplier. The action \eqref{eq:actionfreespin1ontwistor} is invariant under $\Acal\rightarrow \Acal+\bar{\pl}\xi$ and $\Bcal\rightarrow \Bcal+\bar{\pl}\chi$. The pullback $\pi_1^{-1}$ tells us that we should forget about the incident relations and express everything in terms of $\lambda$ and $x$. The measure \eqref{eq:measurePT} becomes
\begin{align}\label{eq:measurePS}
    D^3Z=\lambda_{\alpha}\lambda_{\beta}\langle \lambda d\lambda\rangle \wedge dx^{\alpha\alpha'}\wedge dx^{\beta}_{\ \alpha'}\,.
\end{align}
The integrand in \eqref{eq:actionfreespin1ontwistor} is a $(3,3)$-form on $\PT$ and we need it to have homogeneity 0 in $\lambda$ so that \eqref{eq:actionfreespin1ontwistor} is a well-defined integral. To do so, let us recall that a massless field of helicity $h$ in $\MM^4$ can be represented by a cohomology class, say $\omega$, of Dolbeault cohomology group $H^{0,1}(\PT,\Ocal(2h-2))$ on twistor space. This result can be expressed as 
\begin{equation}\label{eq:twistor1}
    \{\text{massless field on $\MM^4$ of helicity $h$}\}\cong \omega \in H^{0,1}(\PT,\Ocal(2h-2))\,,
\end{equation}
where
\begin{align}
    H^{0,1}(\PT,\Ocal(n)):=\frac{\{\omega \in \Omega^{0,1}(\PT)(n)|\bar{\pl}\omega=0\}}{\{\omega|\omega=\bar{\pl}\beta\}}\,.
\end{align}
Here, $\Omega^{0,1}(\PT)$ is the space of $(0,1)$-forms on $\PT$ of weight $n$, i.e. $f(tZ) = t^nf(Z)$. The usual convention is that a field of negative-helicity has only, say, unprimed indices while a field of positive-helicity has only primed indices. The field equations for free massless fields read
\begin{align}
    \pl^{\alpha\alpha'}\psi_{\alpha(2h)}=0\,,\qquad \pl^{\alpha\alpha'}\tilde{\psi}_{\alpha'(2h)}=0\,.
\end{align}
Asymptotic states formulated in this way turn out to be extremely useful in computing scattering amplitudes, see e.g. \cite{Adamo:2011pv}. However, it is hard to cook up a Lagrangian with objects of different indices. As shown, the action \eqref{eq:actionfreespin1} is only writable with a gauge potential $A^{\alpha,\alpha'}$ which is an object with both primed and unprimed indices. We will need to use one more result in twistor theory. It is proven in \cite{Sparling,Eastwood:1981jy,Hitchin:1980hp} that a gauge potential $\Phi^{\alpha(n),\alpha'}$ that describes a free field of positive helicity and obeys $\pl^{\alpha}_{\ \alpha'}\Phi^{\alpha(n),\alpha'}=0$ is represented by $\Acal\in H^{0,1}(\PT,\Ocal(n-1))$. Another way to phrase this result is
\begin{equation}\label{eq:twistor2} 
   \{\Phi^{\alpha(n),\alpha'}\in \MM^4\,|\, \pl^{\alpha}_{\ \alpha'}\Phi^{\alpha(n),\alpha'}=0\}\cong \Acal\in H^{0,1}(\PT,\Ocal(n-1))\,.
\end{equation}
The equation for $\Phi^{\alpha(n),\alpha'}$ is invariant under $\delta \Phi^{\alpha(n),\alpha'}=\pl^{\alpha\alpha'}\xi^{\alpha(n-1)}$ since $\pl^{\alpha}_{\ \alpha'}\pl^{\alpha\alpha'}\sim \epsilon^{\alpha\alpha}\Box=0$. It is clear that from \eqref{eq:twistor1} and \eqref{eq:twistor2} that the field $\Bcal$ should have weight $-4$ while the gauge potential $\Acal$ should have weight $0$. On $\PS$, it is convenient to work with non-holomorphic coordinates $(x,\lambda)$ where the $(0, 1)$-vector
fields are spanned by
\begin{align}\label{eq:basis}
    \bar{\pl}_0=-\langle \hat{\lambda}\lambda\rangle \lambda_{\alpha}\frac{\pl}{\pl \hat{\lambda}_{\alpha}}\,,\qquad \bar{\pl}_{\alpha'}=-\lambda^{\alpha}\frac{\pl}{\pl x^{\alpha\alpha'}}\,.
\end{align}
Their dual $(0,1)$-forms are
\begin{align}\label{eq:dualbasis}
    \bar{e}^0=\frac{\langle \hat{\lambda}d\hat{\lambda}\rangle}{\langle \hat{\lambda}\lambda\rangle^2}\,,\qquad \bar{e}^{\alpha'}=\frac{\hat{\lambda}_{\alpha}dx^{\alpha\alpha'}}{\langle \hat{\lambda}\lambda\rangle}\,.
\end{align}
It is easy to check that $\bar{\pl}=\bar{e}^0\bar{\pl}_0+\bar{e}^{\alpha'}\bar{\pl}_{\alpha'}$. Varying the action \eqref{eq:actionfreespin1ontwistor} w.r.t. the $\Bcal$ field and expressing $\Acal=\bar{e}^{\alpha'}\Acal_{\alpha'}+\bar{e}^0\Acal_0$ we obtain the following equation:
\begin{align}
    0=\Big(\bar{\pl}_0\Acal_{\alpha'}-\bar{\pl}_{\alpha'}\Acal_0\Big)\bar{e}^0\wedge \bar{e}^{\alpha'}+\bar{\pl}_{\alpha'}\Acal_{\beta'}\bar{e}^{\alpha'}\wedge \bar{e}^{\beta'}\,.
\end{align}
Since $\bar{\pl}_{\alpha'}\Acal_0\,\bar{e}^0\wedge \bar{e}^{\alpha'}=\bar{\pl}(\Acal_0\bar{e}^0)$, we can safely remove $\Acal_{0}$ by a gauge transformation. A different perspective to understand why $\Acal_{0}$ can be set to zero is to use Woodhouse gauge \cite{Woodhouse:1985id} defined with the help of the adjoint differential operator of $\bar{\pl}$, say $\bar{\pl}^*$. By definition, $\bar{\pl}^*\,:\,\Omega^{p,q}(\PT)\mapsto\Omega^{p,q-1}(\PT)$. Due to gauge redundancy, we can restrict
\begin{align}
    \bar{\pl}^*\Acal_{0}|_{\projectivespace^1}=0\,,
\end{align}
which implies that $\Acal_{0}\in \Omega^{0,1}(\projectivespace^1)$. Therefore $\bar{\pl}\Acal_{0}=0$ since $\dim \projectivespace^1=1$. Then, $\Acal_{0}$ is harmonic. According to the Hodge theorem, $\Acal_{0}|_{\projectivespace^1}\in H^{0,1}(\projectivespace^1)$. However, this cohomology group is empty for some specific sheaves $\Ocal$ \cite{Eastwood:1981jy,Woodhouse:1985id}, in particular
\begin{equation}\label{eq:twistor3}
\begin{split}
    H^{0,1}(\projectivespace^1,\Ocal(n))=0\,,\ \text{for} \ n\geq -1\,.
    \end{split}
\end{equation}
Thus, $\Acal_{0}=0$ is a natural gauge condition. The action \eqref{eq:actionfreespin1ontwistor} can be written as
\begin{align}
    S=\int D^3Z \wedge D^3\bar{Z} \Big( \Bcal_{\alpha'}\bar{\pl}_0\Acal^{\alpha'}-\Bcal_0\bar{\pl}_{\alpha'}\Acal^{\alpha'}\Big)\,,
\end{align}
where
\begin{align}
    D^3\bar{Z}=\bar{e}^0\wedge [\bar{e}^{\alpha'}\wedge \bar{e}_{\alpha'}]\,,
\end{align}
and we have expressed $\Bcal=\bar{e}^{\alpha'}\Bcal_{\alpha'}+\bar{e}^0\Bcal_0$. By comparing $\Bcal$ with its spacetime dual $\psi_{\alpha\alpha}$ we see that 
\begin{align}
    \Bcal_0\bar{e}^0=3\frac{\langle \hat{\lambda}d\hat{\lambda}\rangle}{\langle \hat{\lambda}\lambda\rangle^4}\psi_{\alpha\alpha}\hat{\lambda}^{\alpha}\hat{\lambda}^{\alpha}\,,
\end{align}
where the prefactor is chosen for convenience. Since $\Bcal^{\alpha'}$ enters the above action as a Lagrangian multiplier, we get the following constraint $\bar{\pl}_0\Acal_{\alpha'}=0$. This means that $\Acal_{\alpha'}$ should be holomorphic in $\lambda$. From \eqref{eq:twistor2}, we get $\Acal_{\alpha'}=\lambda^{\alpha} A_{\alpha,\alpha'}$. 
Eventually, we find that \cite{Boels:2006ir,Mason:2007ct,Adamo:2011pv}
\begin{align}\label{eq:actionspin1PS}
    \Scal= -\int_{\MM^4} d^4x \int_{\projectivespace^1}\frac{\langle \lambda d\lambda\rangle \wedge \langle \hat{\lambda}d\hat{\lambda} \rangle}{\langle \hat{\lambda}\lambda\rangle^2}\, 3\psi_{\alpha\alpha}\frac{\hat{\lambda}^{\alpha}\hat{\lambda}^{\alpha}\lambda_{\beta}\lambda_{\beta}}{\langle\hat{\lambda}\lambda\rangle^2}\pl^{\beta}_{\ \alpha'}A^{\beta,\alpha'}\,.
\end{align}
The above integrand is of weight zero in both $\lambda$ and $\hat{\lambda}$, and hence it is a well-defined integral. Moreover, the integral over $\projectivespace^1$ in \eqref{eq:actionspin1PS} is a subcase of the following integral \cite{Woodhouse:1985id,Boels:2006ir}
\begin{align}\label{eq:bridge}
    \int_{\projectivespace^1}\frac{\langle \lambda d\lambda\rangle \wedge \langle \hat{\lambda}d\hat{\lambda} \rangle}{\langle \hat{\lambda}\lambda\rangle^{2+m}}\, S_{\alpha(m)}T^{\beta(n)}\LaTeXunderbrace{\hat{\lambda}^{\alpha}...\hat{\lambda}^{\alpha}}_{\text{m times}}\ \LaTeXunderbrace{\lambda_{\beta}...\lambda_{\beta}}_{\text{n times}}\delta_{m,n
    }=-\frac{2\pi i}{m+1}S_{\alpha(m)}T^{\alpha(m)}\,.
\end{align}
Notice that
\begin{align}\label{eq:topform}
    K=\frac{\langle \lambda d\lambda\rangle\wedge\langle \hat{\lambda}d\hat{\lambda}\rangle}{\langle \hat{\lambda}\lambda\rangle^2 }
\end{align}
is nothing but the top form on $\projectivespace^1$. The easiest way to compute the above integral is to plug in local coordinates of $\projectivespace^1$, see appendix \ref{app:twistor}. Using \eqref{eq:bridge}, we see that the spacetime action counterpart of \eqref{eq:actionfreespin1ontwistor} is exactly $\eqref{eq:actionfreespin1}$ up to some prefactor that we can ignore. 

\paragraph{Deformation of the complex structure and SDYM.} The simplest type of interaction from the free action \eqref{eq:actionfreespin1ontwistor} is generated by deforming the complex structure $\bar{\pl}\rightarrow \bar{D}=\bar{\pl}+\Acal$ with $\Acal$ being the matrix-valued $(0,1)$-form connection on $\PT$. The operator $\bar{D}$ is called a covariant almost complex structure, and the twistor action \eqref{eq:actionfreespin1ontwistor} after the deformation reads \cite{Mason:2005zm}
\begin{align}
    \Scal=\int D^3Z\, \Tr\Big[\Bcal \wedge \Big(\bar{\pl}\Acal +\Acal\wedge \Acal\Big)\Big]\,.
\end{align}
This action is invariant under 
\begin{align}\label{eq:gaugetransformBF}
    \delta\Acal= \bar{\pl}\xi +[\Acal, \xi]\,,\qquad \delta \Bcal=\bar{D}\chi+ [\Bcal,\xi]\,,
\end{align}
for smooth sections $\xi \in \Ocal$, and $\chi\in \Ocal(-4)$. One can show that this results in the  Chalmers-Siegel action for SDYM \cite{Chalmers:1996rq} in spacetime by applying the Penrose transform described above \cite{Mason:2005zm,Boels:2006ir,Adamo:2011pv}. We simply quote the final result
\begin{align}\label{eq:actionSiegel}
    S=\int d^4x\, \Tr\Big[\psi_{\alpha\alpha}\pl^{\alpha}_{\  \alpha'}A^{\alpha,\alpha'}\Big]+\int d^4x\, \Tr\Big(\psi_{\alpha\alpha}[A^{\alpha,}_{\ \ \alpha'},A^{\alpha,\alpha'}]\Big)\,.
\end{align}
This action is invariant under the usual Yang-Mills transformations 
\begin{align}
    \delta A^{\alpha,\alpha'}=\pl^{\alpha\alpha'}\xi + [A^{\alpha,\alpha'},\xi]\,,\qquad \delta \psi_{\alpha\alpha}=[\psi_{\alpha\alpha},\xi]\,.
\end{align}
The equation of motion for $A^{\alpha,\alpha'}$ reads
\begin{align}\label{eq:EOMSDYM}
   F^{\alpha\alpha}= \pl^{\alpha}_{\ \alpha'}A^{\alpha,\alpha'}+[A^{\alpha,}_{\ \ \alpha'},A^{\alpha,\alpha'}]=0\,.
\end{align}
\paragraph{Light-cone projection.} Although computations in the light-cone gauge are not aesthetically pleasing at first sight, it turns out to be very useful. Indeed, the light-cone approach opens a direct pathway to work with physical degrees of freedom (avoiding all the gauge redundancy when working with massless fields). For the case of SDYM, following \cite{Chalmers:1996rq}, one can impose $A^{0,0'}=0$ and solve for other components of $A^{\alpha,\alpha'}$ from \eqref{eq:EOMSDYM}. As a result, the two non-vanishing components of $A^{\alpha,\alpha'}$ are:\footnote{See our conventions in appendix \ref{app:convention}.}
\begin{align}
    A^{1,0'}\equiv\Phi_{+1}\,,\qquad A^{1,1'}\equiv \Xi_{+1}=\frac{\bar{\pl}}{\pl^+}\Phi_{+1}\,,
\end{align}
where $\Phi$ is the physical component of $A^{\alpha,\alpha'}$ while $\Xi$ is an auxiliary field. Then, the equation $F^{11}=0$ results in (it is important to remember that $\Phi$ are matrix-valued fields)
\begin{align}
     -\frac{1}{2}\Box\Phi_{+1}+\pl^+\Phi_{+1}\bar{\pl}\Phi_{+1} -\bar{\pl}\Phi_{+1}\pl^+\Phi_{+1}=0\,.
\end{align}
Inside the field $\psi_{\alpha\alpha}$, the physical component is $\psi^{00}=\pl^+\Phi_{-1}$. Hence, we recover the Chalmers-Siegel action \cite{Chalmers:1996rq} in the light-cone gauge:
\begin{align}
    S=\frac{1}{2}\int d^4x\, \Tr[\Phi_{-1} \Box \Phi_{+1}]-\int d^4x\,\Tr\Big[ \Phi_{-1}(\bar{\pl}\Phi_{+1}\pl^+\Phi_{+1}-\pl^+\Phi_{+1}\bar{\pl}\Phi_{+1})\Big]\,.
\end{align}
Having one \textit{spatial} derivative, i.e. the $\bar{\pl}$, in the interaction, the SDYM model lies inside the class of \textit{one-derivative} chiral theories that is shown to be integrable in \cite{Ponomarev:2017nrr}.\footnote{We note that not only the class of one-derivative theories but all chiral theories can be written down in a form that mimics Chalmers-Siegel action in the light-cone gauge. One can further rewrite the equations of motion of the chiral theories as 2d sigma model systems and show that they are integrable because there is an infinite-dimensional symmetry associated with many conserved non-local currents, see more details in \cite{Ponomarev:2017nrr}.} In momentum space, the above action reads
\begin{align}
    S=-\frac{1}{2}\int d^4\pvec\, \Tr[\Phi_{-1}\Phi_{+1}]\pvec^2+\int d^4\pvec\, \PPb\,\Tr[\Phi_{-1}\Phi_{+1}\Phi_{+1}]\,,
\end{align}
where $\pvec:=(\beta,p^-,p,\bar{p})$ and $\PPb_{ij}:=\bar{p}_i\beta_j-\bar{p}_j\beta_i$ for $\pvec_i$ being the momenta of the field $\Phi_i$. For more on the light-cone gauge approach, see \cite{Metsaev:1991mt,Metsaev:1991nb,Ponomarev:2016lrm,Ponomarev:2017nrr}.

\section{Twistor actions from the inverse Penrose transform}\label{sec:2}
After reviewing all the required elements to construct the map between the twistor space and spacetime, we can now present the inverse Penrose transform. The input is a covariant action on spacetime that can be written as self-dual part plus non-self-dual part with \eqref{eq:twistor1} and \eqref{eq:twistor2} being the main criteria to construct the dual twistor action. We examine the well-known cases of SDYM and the full Yang-Mills in this section. The first step is to insert the integral over the top-form of $\projectivespace^1$ to \textit{lift} spacetime actions to the corresponding space $\PS$. Then, one can project data from $\PS$ to $\PT$ via the projection $\pi_1$ as shown in \eqref{doublefibration}. 
\subsection{Twistor construction for SDYM}\label{sec:SDYM}
The shorthand for the SDYM \eqref{eq:actionSiegel} is
\begin{align}\label{eq:actionBFspacetimeSDYM}
    S=\int d^4x\, \Tr\big(\psi_{\alpha\alpha}F^{\alpha\alpha}\big)\,.
\end{align}
Putting back the integration measure $K$ on $\projectivespace^1$, we can write
\begin{align}
    d^4x\frac{\langle \lambda d\lambda\rangle \wedge \langle \hat{\lambda}d\hat{\lambda}\rangle }{\langle \hat{\lambda}\lambda\rangle^2}=D^3Z\wedge D^3\bar{Z}\,.
\end{align}
From \eqref{eq:twistor1}, we know that $\psi_{\alpha\alpha}$ should be expressed as a $(0,1)$-form of weight $-4$ on the twistor space, while from \eqref{eq:twistor2}, $F^{\alpha\alpha}$ should become a curvature $(0,2)$-form of weight $0$. Using \eqref{eq:bridge}, the corresponding action of \eqref{eq:actionBFspacetimeSDYM} in twistor space reads (we once again ignore the constant prefactor)
\begin{align}
    \Scal=-\int d^4x\int_{\projectivespace^1}\langle \lambda d\lambda\rangle\, \Tr\Big[3\frac{\langle \hat{\lambda}d\hat{\lambda}\rangle}{\langle \hat{\lambda}\lambda\rangle^4}\psi_{\beta\beta}\hat{\lambda}^{\beta}\hat{\lambda}^{\beta}\, F^{\alpha\alpha}\lambda_{\alpha}\lambda_{\alpha}\Big]\,.
\end{align}
By imposing the incident relations and using the definition \eqref{eq:basis}, we get
\begin{align}\label{eq:BFactionforSDYM}
    \Scal=\int_{\PT}D^3Z \,\Tr\Big(\Bcal\wedge \Fcal\Big)\,,
\end{align}
where $\Bcal$ is a $(0,1)$-form of weight $-4$ in $\lambda$ and $\Fcal$ is a curvature $(0,2)$-form of weight 0:
\begin{subequations}
\begin{align}
    \Bcal &= 3\frac{\langle \hat{\lambda}
d\hat{\lambda}\rangle }{\langle \hat{\lambda}\lambda\rangle^4 }\psi_{\alpha\alpha}\hat{\lambda}^{\alpha}\hat{\lambda}^{\alpha}=\Bcal_0\bar{e}^0\,,\\
\Fcal&=-\lambda_{\alpha}\lambda_{\alpha}F^{\alpha\alpha}[\bar{e}^{\gamma'}\wedge \bar{e}_{\gamma'}]=\Big(\bar{\pl}_{\alpha'}\Acal_{\beta'}+[\Acal_{\alpha'},\Acal_{\beta'}]\Big)\bar{e}^{\alpha'}\wedge \bar{e}^{\beta'}\,.
\end{align}
\end{subequations}
Notice that we only recovered \textit{half} of the original twistor action for SDYM using the inverse map up to this point since the twistor fields are in the Woodhouse gauge. The action \eqref{eq:BFactionforSDYM} is invariant under the gauge transformations \eqref{eq:gaugetransformBF}, and therefore we can recover the missing components $\Bcal_{\alpha'}$ of the Lagrangian multiplier fields, and $\Acal_0$ of the $(0,1)$-form gauge potential on twistor space. We conclude that the twistor action of SDYM is the action \eqref{eq:BFactionforSDYM}. In addition, using the inverse Penrose transform, one can also show that the twistor action for $\Ncal=4$ SDYM \cite{Chalmers:1996rq,Siegel:1992xp} is dual to the Chern-Simons action on the super twistor space $\projectivespace^{3|4}$ \cite{Witten:2003nn,Boels:2006ir}.

\subsection{Twistor construction for Yang-Mills}
To get the full Yang-Mills theory, one can add a deformation to the SDYM action as a perturbation around the self-dual sector \cite{Mason:2005zm}
\begin{align}\label{eq:actionYMspacetime}
    S=\int d^4x\, \Tr[\psi_{\alpha\alpha}F^{\alpha\alpha}]-\frac{g}{2}\int d^4x\,\Tr[ \psi_{\alpha\alpha}\psi^{\alpha\alpha}]\,,
\end{align}
where $g$ is a coupling. One can show that this action is equivalent to Yang-Mills theory. Indeed, by varying \eqref{eq:actionYMspacetime} with respect to $\psi_{\alpha\alpha}$ one obtains $F^{\alpha\alpha}=g\psi^{\alpha\alpha}$. Hence,
\begin{align}
    S=\frac{1}{2g}\int d^4x\, \Tr[F_{\alpha\alpha}F^{\alpha\alpha}]\,.
\end{align}
The above action depends only on the self-dual field strength of the gauge field and is equivalent to the Yang-Mills action. Next, recall that the pullback of $\psi$ is a $(0,1)$-form on the twistor space $\PT$. However, since $\psi$s (inside the $\psi^2$ term) should have the same form-degree with the $\Bcal$ field after being lifted to the twistor space, we can consider a new space that is the fiber-wise product of $\PT$ with itself over $\MM$ with fiber $\projectivespace^1 \times  \projectivespace^1$, i.e. $\PT\times_{\MM} \PT=\MM^4\times \projectivespace^1\times \projectivespace^1$. The coordinates in this space are $(Z_1,Z_2)$ such that $\pi_2(Z_1)=\pi_2(Z_2)=x^{\alpha\alpha'}$ to maintain locality in spacetime.\footnote{This is a special feature of the twistor space. Since it plays the role of an auxiliary space, it is harmless to have non-local interactions on twistor space. One just needs to make sure that the interactions in spacetime are local after the Penrose transform.} In terms of components, these coordinates read
\begin{align}
    Z_1=(\lambda_{1\alpha},\mu_1^{\alpha'}=x^{\alpha\alpha'}\lambda_{1\alpha})\,,\qquad Z_2=(\lambda_{2\alpha},\mu_2^{\alpha'}=x^{\alpha\alpha'}\lambda_{2\alpha})\,.
\end{align}
Let us consider a pair of twistors $\hat{Z}_{1,2}$ that are defined with respect to the fiber-wise product $\times_{\MM}$ and Euclidean signature conjugation operation ' $\hat{}$ ' in \eqref{eq:hatoperator}:
\begin{align}
    \hat{Z}_1=(\hat{\lambda}_{1\alpha},\hat{\mu}_1^{\alpha'}=x^{\alpha\alpha'}\hat{\lambda}_{1\alpha})\,,\qquad \hat{Z}_2=(\hat{\lambda}_{2\alpha},\hat{\mu}_2^{\alpha'}=x^{\alpha\alpha'}\hat{\lambda}_{2\alpha})\,.
\end{align}
The twistor dual of $g\int \Tr[\psi^2]$ is obtained by inserting integrals over $\projectivespace^1$ with $K_1$ and $K_2$ (recall that $K$ is the top-form on $\projectivespace^1$ defined in \eqref{eq:topform}) being the measures 
\begin{align}\label{eq:psi2twistor}
    -\frac{g}{2}\int D^3Z_1 \wedge D^3Z_2 \,\Tr\big[\Bcal_1\Bcal_2\big]\,,
\end{align}
where 
\begin{align}
    \Bcal_i=3\langle \hat{\lambda}_id\hat{\lambda}_i\rangle \frac{\psi_{\alpha\alpha}\hat{\lambda}_i^{\alpha}\hat{\lambda}_i^{\alpha}}{\langle \hat{\lambda}_i\lambda_i\rangle^4}\,.
\end{align}
Observe that \eqref{eq:psi2twistor} is a two-point integral on twistor space where $\Bcal_i$ are cohomology classes of $H^{0,1}(\PT,\Ocal(-4))$ as before. The twistor action of Yang-Mills theory is therefore
\begin{align}\label{eq:YMtwistoraction}
    \Scal=\int D^3Z\, \Tr[\Bcal\Fcal]-\frac{g}{2}\int D^3Z_1\wedge  D^3Z_2\,\Tr[\Bcal_1\Bcal_2]\,.
\end{align}
Up to this point, we have restricted ourselves in the Woodhouse gauge because it is convenient to perform the (inverse) Penrose transform. However, one can switch from Woodhouse gauge to axial gauge \cite{Cachazo:2004kj} to study scattering amplitudes from twistor space, see \cite{Boels:2007qn, Adamo:2011pv} for the cases of (supersymmetric) Yang-Mills theory. We expect that the twistor construction via the inverse Penrose transform should work for more general cases. In particular, whenever we can decompose a one-derivative spacetime action into a self-dual part and a non-self-dual part, we should be able to lift the spacetime action to twistor space. The correspondence twistor space for the non-self-dual part of the action is $X=\MM^4\times (\projectivespace^1)^{\otimes n}$ where $n$ is the number of fields with non-positive helicities in each vertex, see \cite{Boels:2006ir} for the example of $\Ncal=4$ SYM.

\section{Twistor actions of free scalar field and free higher-spin fields}\label{sec:3}
In this section, we first show how one can construct a twistor action for the scalar field. Then, we show how to generalize the twistor action of a free spin-1 field to higher-spin fields by considering \textit{generalized connections} and \textit{generalized Lagrangian multiplier twistor fields} on twistor space.

\subsection{The scalar field}
Having the scalar field in the spectrum is essential for generic higher spin theories to be consistent in spacetime. For that reason, let us start this subsection by constructing the twistor action of a free scalar field. Recall that the free action for a scalar field is $-\frac{1}{2}(\pl\phi)^2$. Let us introduce an auxiliary field  $\vartheta$, and write the free action for a scalar field as
\begin{align}\label{eq:scalaraux}
    \int d^4x \Big(\phi \pl^{\alpha}_{\ \alpha'}\vartheta_{\alpha}^{\ \alpha'}+\frac{\vartheta^{\alpha}_{\ \alpha'}\vartheta_{\alpha}^{\ \alpha'}}{2}\Big)\,.
\end{align}
It is easy to check that this action is equivalent to the free scalar action after integrating out the auxiliary field $\vartheta$. Observes that \eqref{eq:scalaraux} is a one-derivative action, and hence feasible to be lifted to twistor space by the inverse Penrose transform. The dual twistor action of \eqref{eq:scalaraux} reads
\begin{align}
    -\int D^3Z\wedge  D^3\bar{Z} \Big(\phi\bar{\pl}_{\alpha'}\tilde{\vartheta}^{\alpha'}-\frac{1}{2}\tilde{\vartheta}^{\alpha'}\lambda_{\alpha}\vartheta^{\alpha}_{\ \alpha'}\Big)\,,
\end{align}
where $\tilde{\vartheta}^{\alpha'}=\frac{1}{\langle \hat{\lambda}\lambda\rangle}\hat{\lambda}^{\alpha}\vartheta_{\alpha}^{\ \alpha'}$. Using the basis \eqref{eq:basis}, we can write the above as
\begin{align}\label{eq:actionintermediatescalar}
 -\int D^3Z\wedge  D^3\bar{Z}\Big(\phi\bar{\pl}_{\alpha'}\tilde{\vartheta}^{\alpha'}+\frac{1}{2}\tilde{\vartheta}_{\alpha'}\bar{\pl}_0\tilde{\vartheta}^{\alpha'}\Big)\,.
\end{align}
The action \eqref{eq:actionintermediatescalar} can be further reduced to
\begin{align}\label{eq:twistorscalaraction}
    \int D^3Z\, \varphi \bar{\pl}\varphi\,,
\end{align}
where 
\begin{align}
    \varphi=\phi\bar{e}^0+\tilde{\vartheta}_{\alpha'}\bar{e}^{\alpha'}\,.
\end{align}
Therefore, $\varphi\in H^{0,1}(\PT,\Ocal(-2))$, which follows from \eqref{eq:twistor1}. We note that the above twistor construction of the scalar field does not require supersymmetry. A nice feature of \eqref{eq:twistorscalaraction} is that it contains only one derivative instead of two like in spacetime.

\subsection{Free higher spin fields}
Moving to free higher spin fields, there is a simple action that satisfies both \eqref{eq:twistor1} and \eqref{eq:twistor2}, and is a straightforward generalization of \eqref{eq:actionfreespin1}. It reads
\begin{align}\label{eq:actionfreeHS}
    S=\int d^4x\,\psi_{\alpha(2s)}\pl^{\alpha}_{\ \alpha'}\Phi^{\alpha(2s-1),\alpha'}\,.
\end{align}
The above action is invariant under $\delta \Phi^{\alpha(2s-1),\alpha'}=\pl^{\alpha\alpha'}\xi^{\alpha(2s-2)}$. One can go back to the usual second order action by considering
\begin{align}
    S=\int d^4x\, \psi_{\alpha(2s)}\pl^{\alpha}_{\ \alpha'}\Phi^{\alpha(2s-1),\alpha'}-\frac{1}{2}\int d^4x \, \psi_{\alpha(2s)}\psi^{\alpha(2s)}\sim \int d^4x \, \Big(\pl^{\alpha}_{\ \alpha'}\Phi^{\alpha(2s-1),\alpha'}\Big)^2\,.
\end{align}
The EOM resulting from the above action is gauge invariant and describes both helicities. Note that \eqref{eq:actionfreeHS} is valid for any spin, including half-integer ones. It is the advantage of using spinorial indices instead of Lorentz ones. The above formulation is in contrast to the usual expression of free higher spin fields in terms of Fronsdal fields which is quadratic in derivatives for free bosonic fields and linear in derivatives for free fermionic fields. We will employ the first order action \eqref{eq:actionfreeHS} because it is easier to work with when we consider interactions between higher-spin fields. 

In addition, as discussed in \cite{Krasnov:2021nsq}, fields that are written in terms of pure unprimed indices or have only one primed indices are said to live in \textit{maximally unbalanced representations} of the Lorentz group $S(m,n)$ in 4d. It turns out that this description of higher-spin fields is more useful than the usual formulation using Fronsdal fields.\footnote{One of the main motivations to use the above expression is that higher-spin theories written in terms of Fronsdal fields contain non-local vertices starting from the quartic \cite{Bekaert:2015tva}. This results in a severe No-Go obstruction for the existence of higher-spin theories with interactions, see e.g. \cite{Ponomarev:2017nrr,Bekaert:2010hp,Fotopoulos:2010ay,Boulanger:2015ova,Roiban:2017iqg,Bekaert:2015tva,Sleight:2017pcz}.} Indeed, one can write down Lorentz covariant actions that are invariant off-shell for HS-SDYM, self-dual gravity, and its higher spin extensions \cite{Krasnov:2016emc,Krasnov:2021nsq} using the above representation. Moreover, higher spin fields described by maximally unbalanced representations can live on not only maximally symmetric backgrounds such as (A)dS but also on more general backgrounds that are self-dual. Here, by self-dual backgrounds, we mean all backgrounds whose half of the Weyl tensor vanishes. The twistor action for a free spin-$s$ field reads
\begin{align}\label{eq:freeHStwistor}
    \int D^3Z\,\Bcal\bar{\pl}\omega\,,
\end{align}
where
\begin{subequations}
\begin{align}
    \omega&=\omega_{\alpha'}\bar{e}^{\alpha'}+\omega_0\bar{e}^0=\LaTeXunderbrace{\lambda^{\alpha}\cdots \lambda^{\alpha}}_{2s-1\  \text{times}} \Phi_{\alpha(2s-1),\alpha'}\bar{e}^{\alpha'}+\omega_0^s\bar{e}^0\,,\\
    \Bcal&=\Bcal_0\bar{e}^0+\Bcal_{\alpha'}\bar{e}^{
\alpha'}=(2s+1)\LaTeXunderbrace{\hat{\lambda}^{\alpha}\cdots \hat{\lambda}^{\alpha}}_{2s\  \text{times}}\psi_{\alpha(2s)}\frac{\langle \hat{\lambda}
d\hat{\lambda}\rangle }{\langle \hat{\lambda}\lambda\rangle^{2s+2} }+\Bcal_{\alpha'}^s\bar{e}^{\alpha'}\,.
\end{align}
\end{subequations}
We will refer $\omega$ as \textit{generalized connection} and $\Bcal$ as \textit{generalized Lagrangian multiplier twistor field}. The free twistor action for a free spin-$s$ field \eqref{eq:freeHStwistor} is invariant under
\begin{align}\label{eq:gaugetransformfreeHSonPT}
    \delta\omega=\bar{\pl}\xi\,,\qquad \delta\Bcal=\bar{\pl}\chi\,,
\end{align}
for smooth sections $\xi\in \Ocal(2s-2)$, and $\chi\in \Ocal(-2s-2)$. We once again note that to go to spacetime action \eqref{eq:actionfreeHS} from \eqref{eq:freeHStwistor}, it is most convenient to use Woodhouse gauge.

\section{Twistor constructions for HS-SDYM and HS-YM}\label{sec:4}
In this section, we construct the twistor action of HS-SDYM \cite{Krasnov:2021nsq} via the inverse Penrose transform. By adding a small perturbation to HS-SDYM, we obtain a higher-spin extension action for Yang-Mills which we denote HS-YM for short. The dual twistor action of HS-SDYM has a similar form to $\Bcal\Fcal$-action. Therefore, from the twistor point of view, HS-SDYM should be integrable and quasi-topological in spacetime. 
\subsection{Twistor action of HS-SDYM}
The action for HS-SDYM \cite{Krasnov:2021nsq} (or one-derivative chiral higher-spin theory) reads
\begin{align}\label{eq:HSSDYMspacetime}
    S=\sum_{s}\int \Tr[\psi_{\alpha(2s)}\pl^{\alpha}_{\ \alpha'}\Phi^{\alpha(2s-1),\alpha'}]+\sum_{m+n=2s}\Tr\Big(\psi_{\alpha(m+n)}[\Phi^{\alpha(m),}_{\quad \quad \alpha'},\Phi^{\alpha(n),\alpha'}]\Big)\,.
\end{align}
This action is invariant under the higher-spin extension of Yang-Mills gauge transformation
\begin{align}\label{eq:HSgaugetransform}
    \delta \Phi^{\alpha(2s-1),\alpha'}=\pl^{\alpha\alpha'}\xi^{\alpha(2s-2)}+[\Phi,\xi]^{\alpha(2s-1),\alpha'}\,,\qquad \delta\psi_{\alpha(2s)}=[\psi,\xi]_{\alpha(2s)}\,.
\end{align}
The light-cone action for HS-SDYM is closely related to the Chalmers-Siegel action
\begin{align}
   S=-\frac{1}{2}\sum_{s\geq 1}\int d^4\pvec\, \Tr[ \Phi_{-s}\Phi_{+s}]\pvec^2 +\sum_{s_1,s_2}\int d^4\pvec \, a_{s_1,s_2}\PPb\,\Tr\big[\Phi_{-(s_1+s_2-1)}\Phi_{+s_1}\Phi_{+s_2}\big]\,.
\end{align}
\paragraph{Twistor action.} For the case of HS-SDYM, higher-spin symmetry, i.e. the interactions between higher-spin fields, is encoded by the sum over the unprimed indices in \eqref{eq:HSSDYMspacetime}. This information can be realized by considering the following twistor fields
\begin{align}\label{eq:HSconnection}
     \omega&\equiv\bar{e}^{\alpha'}\omega_{\alpha'}+\bar{e}^0\omega_0=\sum_s\big(\omega_{\alpha'}^s\bar{e}^{\alpha'}+\omega_0^s\bar{e}^0\big)=\sum_s \LaTeXunderbrace{\lambda^{\alpha}\cdots \lambda^{\alpha}}_{2s-1\  \text{times}} \Phi_{\alpha(2s-1),\alpha'}\bar{e}^{\alpha'}+\sum_s\omega_0^s\bar{e}^0\,,
\end{align}
and
\begin{align}\label{eq:HSmultiplier}
     \Bcal&\equiv\bar{e}^0\Bcal_0+
    \bar{e}^{\alpha'}\Bcal_{\alpha'}=\sum_s\big(\Bcal_0^s\bar{e}^0+\Bcal_{\alpha'}^s\bar{e}^{
\alpha'}\big)=\sum_s(2s+1)\,\LaTeXunderbrace{\hat{\lambda}^{\alpha}\cdots \hat{\lambda}^{\alpha}}_{2s\  \text{times}}\psi_{\alpha(2s)}\frac{\langle \hat{\lambda}
d\hat{\lambda}\rangle }{\langle \hat{\lambda}\lambda\rangle^{2s+2} }+\sum_s\Bcal_{\alpha'}^s\bar{e}^{\alpha'}\,.
\end{align}
The dual twistor action of \eqref{eq:HSSDYMspacetime} is therefore
\begin{align}\label{eq:BFHStwistor}
    \Scal=\int D^3Z \, \Tr[\Bcal(\bar{\pl}\omega+\omega\wedge \omega)]\,.
\end{align}
One can easily check that the action \eqref{eq:BFHStwistor} reduces to \eqref{eq:HSSDYMspacetime} after the Penrose transform in the Woodhouse gauge. On twistor space, when interactions between higher-spin fields are taken into account, the generalized connection \eqref{eq:HSconnection} becomes cohomology class of
\begin{align}\label{eq:generalizedomega}
    \omega\in \bigoplus_s H^{0,1}(\PT,\Ocal(2s-2))=H^{0,1}(\PT,\bigoplus_s\Ocal(2s-2))\,.
\end{align}
Similarly, the generalized Lagrangian multiplier field $\Bcal$ are now cohomology class of
\begin{align}\label{eq:generalizedB}
    \Bcal\in \bigoplus_s H^{0,1}(\PT,\Ocal(-2s-2))=H^{0,1}(\PT,\bigoplus_s \Ocal(-2s-2))\,.
\end{align}
\paragraph{Uniqueness.} Now, let us prove that the above action is unique. Similar to the case of the twistor action for SDYM, consider the following deformation of the complex structure $\bar{\pl}$ on twistor space:
\begin{align}\label{eq:deformedDHSPT}
    \bar{D}=\bar{\pl}+\omega\,.
\end{align}
where $\omega=\bar{e}^{\alpha'}\omega_{\alpha'}+\bar{e}^0\omega_0=\sum_s\big(\omega_{\alpha'}^s\bar{e}^{\alpha'}+\omega_0^s\bar{e}^0\big)$ on $\PT\cong \PS$. Using the basis \eqref{eq:basis} and \eqref{eq:dualbasis}, one can decompose $\bar{D}$ into
\begin{align}
    \bar{D}\equiv \bar{e}^0\bar{D}_0+\bar{e}^{\alpha'}\bar{D}_{\alpha'}=\bar{e}^0\big(\bar{\pl}_0+\omega_0\big)+\bar{e}^{\alpha'}\big(\bar{\pl}_{\alpha'}+\omega_{\alpha'}\big)\,.
\end{align}
The integrability condition $\Fcal=\bar{D}^2=0$ reads
\begin{align}\label{eq:twistorintegrability}
   \Fcal=\bar{D}\omega=\bar{\pl}\omega+\omega\wedge \omega= \Big(\bar{\pl}_0\omega_{\alpha'}-\bar{D}_{\alpha'}\omega_0\Big)\bar{e}^0\wedge \bar{e}^{\alpha'}+\bar{D}_{\alpha'}\omega_{\beta'}\,\bar{e}^{\alpha'}\wedge \bar{e}^{\beta'}=0\,,
\end{align}
where $\bar{D}_{\alpha'}\bullet=\bar{\pl}_{\alpha'}+[\omega_{\alpha'},\bullet]$. Note that $\bar{D}_{\alpha'}\omega_0\, \bar{e}^0\wedge \bar{e}^{\alpha'}=\bar{D}(\omega_0\bar{e}^0)$ and therefore could be removed by the \textit{generalized} gauge transformation $\omega\rightarrow \omega+\bar{D}\xi$ for some smooth sections $\xi$ (see below). Therefore, we can always restrict ourselves to the Woodhouse gauge, where we set $\omega_0$ to zero, starting from \eqref{eq:twistorintegrability}. In terms of components, the integrability condition \eqref{eq:twistorintegrability} is equivalent to
\begin{subequations}\label{eq:systemofintegrability}
\begin{align}
    0&=\sum_s\bar{\pl}_0\omega_{\alpha'}^s\,,\\
    0&=\sum_s\bar{\pl}_{\alpha'}\omega_{\beta'}^s+\sum_{m+n=s}[\omega_{\alpha'}^m,\omega_{\beta'}^n]\,,
\end{align}
\end{subequations}
The first equation in \eqref{eq:systemofintegrability} tells us that $\omega_{\alpha'}^s$ should be holomorphic in $\lambda$. In addition, treating \eqref{eq:twistor2} as the constraint on positive-helicity fields, then 
\begin{align}
    \omega_{\alpha'}^s=\LaTeXunderbrace{\lambda^{\alpha}\cdots \lambda^{\alpha}}_{2s-1\  \text{times}} \Phi_{\alpha(2s-1),\alpha'}\,,
\end{align}
as expected. Therefore, we interpret that $\omega \in H^{0,1}(\PT,\bigoplus_s \Ocal(2s-2))$. From this, the second equation in \eqref{eq:systemofintegrability} gives
\begin{equation}
\begin{split}\label{eq:selfdualcondition}
    0&=\epsilon_{\alpha'\beta'}\sum_s\LaTeXunderbrace{\lambda^{\alpha}\cdots \lambda^{\alpha}}_{2s\  \text{times}}\Big(\pl_{\alpha \gamma'} \Phi_{\alpha(2s-1),}^{\qquad \quad \gamma'}+\sum_{m+n=2s}[\Phi_{\alpha(m),\gamma'},\Phi_{\alpha(n),}^{\quad \ \  \gamma'}]\Big)\,\\
    &=\epsilon_{\alpha'\beta'}\sum_s\LaTeXunderbrace{\lambda^{\alpha}\cdots \lambda^{\alpha}}_{2s\  \text{times}}F_{\alpha(2s)}\,.
    \end{split}
\end{equation}
The above equation can only be satisfied if $F_{\alpha(2s)}=0$. 
Notice, that $F_{\alpha(2s)}$ is the same with the equation of motion for $\Phi$ that can be derived from the spacetime action \eqref{eq:HSSDYMspacetime}. Hence, we have shown that the generalized covariant almost complex structure $\bar{D}$ defined in \eqref{eq:deformedDHSPT} obeying $\Fcal=0$ leads to HS-SDYM in spacetime. By introducing a Lagrangian multiplier field $\Bcal\in \Omega^{0,1}(\PT)$, one can write down a twistor action for the system \eqref{eq:twistorintegrability} as \eqref{eq:BFHStwistor}. The equation for the generalized Lagrangian multiplier field $\Bcal$ reads
\begin{align}
    \bar{D}\Bcal=(\bar{\pl}+\omega)\Bcal=0\,.
\end{align}
The above equation is invariant under $\delta \Bcal\rightarrow \Bcal+ \bar{D}\chi$. Therefore, we can understand $\Bcal$ as cohomology classes on $\PT$. By virtue of \eqref{eq:twistor1}, \eqref{eq:bridge} and \eqref{eq:selfdualcondition}, we conclude that $\Bcal\in H^{0,1}(\PT,\bigoplus_s\Ocal(-2s-2))$. Combining all of the above, we see that the twistor action \eqref{eq:BFHStwistor} is invariant under
\begin{align}\label{eq:twistorHSgaugetransformation}
    \delta\omega=\bar{\pl}\xi+[\omega,\xi]\,,\qquad \delta\Bcal=[\Bcal,\xi]+\bar{\pl}\chi+[\omega,\chi]\,,
\end{align}
for generalized smooth sections $\xi\in \bigoplus_s\Ocal(2s-2)$, and $\chi\in \bigoplus_s\Ocal(-2s-2)$. Note that the weight of $[\omega,\xi]$ should match with the weight of $\delta\omega$ for the generalized gauge transformation to work. Finally, let us give a brief comment on the above result. At the linearize level, the equations of $\omega$ and $\Bcal$ reduces to $\bar{\pl}\omega=0$ and $\bar{\pl}\Bcal=0$. Being invariant under the gauge transformation \eqref{eq:gaugetransformfreeHSonPT}, we can interpret $\omega$ and $\Bcal$ as defining elements in the $\bar{\pl}$-cohomology groups $H^{0,1}(\PT,\Ocal(2s-2))$ and $H^{0,1}(\PT,\Ocal(-2s-2))$, respectively. The situation becomes different when we consider interactions between higher-spin fields in spacetime. 
On twistor space, higher-spin symmetry can be represented by the cohomology classes \eqref{eq:generalizedomega} and \eqref{eq:generalizedB} which obey the generalized gauge symmetry \eqref{eq:twistorHSgaugetransformation}. 
\paragraph{UV finiteness.} The action of HS-SDYM has a form similar to the $\Bcal\Fcal$ action. We know that the $\Bcal\Fcal$ action is topological and is one-loop exact on twistor space because the interaction is of $\Bcal \Acal \Acal$ type (the wedge product is implicit here). These properties of the $\Bcal\Fcal$ action should \textit{descend} to spacetime through the Penrose transform. Hence, besides being integrable, HS-SDYM should also be quasi-topological, i.e. almost free, and one-loop exact. Indeed, one can show that the tree-level amplitudes for HS-SDYM vanish and the one-loop result is closely related to the results of \cite{Chalmers:1996rq,Bern:1998sv,Skvortsov:2020gpn}.

\subsection{HS-YM and its twistor action}
Similar to the case of Yang-Mills theory, we can try to perturb the action of HS-SDYM by adding a $\psi^2$ term to see whether we can obtain an action that is a higher-spin extension of the Yang-Mills one. Let us consider the following action
\begin{align}
    S=\sum_m\int d^4x\,\Tr[\psi_{\alpha(2m)}F^{\alpha(2m)}]-\sum_m\frac{g_m}{2}\int \Tr[\psi_{\alpha(2m)}\psi^{\alpha(2m)}]\,.
\end{align}
As in the case of Yang-Mills, one can solve for $\psi$ in terms of $F$ and plug it back in. Then, the action for the higher-spin extension of Yang-Mills reads
\begin{align}\label{eq:actionHSYMspacetime}
    S=\sum_m\frac{1}{2g_m}\int d^4x \, \Tr[F_{\alpha(2m)}F^{\alpha(2m)}]\,.
\end{align}
It is easy to see that this action is gauge invariant under the higher-spin extension of a Yang-Mills gauge transformation \eqref{eq:HSgaugetransform}.\footnote{A detailed investigation of the action \eqref{eq:actionHSYMspacetime} will be presented in a companion paper.} Similarly to the case of Yang-Mills theory, the twistor action of HS-YM reads \begin{align}
    \Scal=\int D^3Z \,\Tr[\Bcal(\bar{\pl}\omega+\omega\wedge \omega)]-\frac{g_m}{2}\int D^3Z_1\wedge D^3Z_2\,\Tr[\Bcal_1\Bcal_2]\,,
\end{align}
where $\Bcal$ and $\omega$ are defined in \eqref{eq:HSconnection} and \eqref{eq:HSmultiplier}. We note that all theories with at most one \textit{spatial} derivative (in the light-cone gauge) in the interaction terms and expanded around their self-dual sectors should be writable in this way. It would be interesting to apply this perturbation technique to the HS-SDGRA and the (parent) chiral HSGRA to see whether we can achieve parity-invariant theories from their self-dual sectors.

\section{Discussion}\label{conclusion}
In this paper, we presented the inverse Penrose transform on flat spacetime for various known examples and some new examples of HS-(SD)YM. The most important takeaway from this paper is that the inverse Penrose transform can help us gain insight into the twistor construction of theories in spacetime. Due to the twistor correspondence, most (if not all) properties of the twistor actions will descend to spacetime via the Penrose transform. We have shown that all self-dual theories considered in this paper are dual to (generalized) $\Bcal\Fcal$ theory on twistor space. Therefore, they are integrable from the twistor perspective. 

The results of the inverse mapping can be summarized as follows. Consider a spacetime action of the one-derivative type in the main text that lives on flat spacetime. Then, the inverse map exists if we can decompose the spacetime action into a self-dual sector plus a sector that contains non-positive-helicity fields. It would be interesting to test this idea for other theories. One of the purposes of studying this inverse mapping is to guess (or construct) the mechanism on twistor space that can generate higher derivative interactions in spacetime. 

As a remark, in this paper, we write down fields in the maximally unbalanced representations of the Lorentz group. This approach turns out to be more natural, compared to the formulations that employ Lorentz indices, towards the goal of classifying all possible vertices \cite{Bengtsson:2014qza,Conde:2016izb} and writing down a consistent off-shell action with higher-spin interactions, see for examples \cite{Ponomarev:2016lrm,Krasnov:2021nsq}. In addition, theories formulated in the maximally unbalanced representations are closely related to what is known as the spinor-helicity formalism that is used extensively in computing scattering amplitudes due to its efficiency. Therefore, having a covariant action written in terms of spinorial indices to start with maybe beneficial since one can employ amplitudes techniques to investigate the theory. It is indeed the case for HS-SDYM and HS-SDGRA \cite{Krasnov:2021nsq}. Since we can write down Lorentz covariant descriptions for these two examples, some of the puzzles in higher spin have been resolved.

It would be compelling to construct (or at least guess) the twistor action for the SDGRA and its higher-spin extension \cite{Krasnov:2021nsq} using the inverse Penrose transform. Originally, the SDGRA action in \cite{Krasnov:2016emc} was derived from what is called the $b\tau$-action on twistor space \cite{Mason:2007ct}:
\begin{align}
    S[b,\tau]=\int_{\PS} b\wedge \tau\wedge d\tau\wedge d\tau\,.
\end{align}
Even though the SDGRA in flat space contains only cubic interactions, its twistor origin is quartic in the fields. Hence, it is hard to immediately predict whether SDGRA is integrable from the twistor perspective. The higher-spin extension of SDGRA also admits a very similar form of the SDGRA action in spacetime. Therefore, with the help of the inverse map, we should be able to obtain the dual twistor actions that contain only cubic interactions for these two theories. It would naturally give a geometrical explanation of why SDGRA and its higher-spin extension are integrable.

The full Yang-Mills is obtained by perturbing around its self-dual sector. The strategy of adding extra terms to the self-dual sector is used in gravity as well. Indeed, by adding an infinite number of terms to the Lagrangian of SDGRA, one recovers GR \cite{Krasnov:2016emc}. We applied this technique to obtain HS-YM from HS-SDYM. It would be interesting to see whether we can discover a large class of parity-invariant higher spin theories with propagating d.o.f in this way. It is then compelling to check whether parity-invariant higher spin theories obtained by deforming the self-dual ones make sense by performing the analysis in the light-cone gauge similar to \cite{Ponomarev:2016lrm}.

There are several proposals for twistor deformations that result in higher-derivative interactions, e.g. deformation of the connection \cite{Haehnel:2016mlb,Adamo:2016ple}, deformation using the Poisson bracket or the star product \cite{Mason:2007ct}, etc. We hope that with the help of twistor theory, one can gain enough intuition to finally write down a covariant action for the chiral HSGRA. Moreover, it would be interesting to find a general approach for deformations on twistor space such that we can have a covariant action in spacetime with higher-derivative interactions.

\section*{Acknowledgements}
I am grateful to Lionel Mason, Dmitry Ponomarev, Atul Sharma, Zhenya Skvortsov and Stefan Theisen constructive discussions. I appreciate Zhenya Skvortsov very useful guidance in the course of collaborating on \cite{Krasnov:2021nsq}. I would like to thank the anonymous JHEP referee for many valuable suggestions and improvements. Moreover, the discussions on twistor theory with Maxim Grigoriev, Lionel Mason, Tristan McLoughlin, Arthur Lipstein, and Ivo Sachs that date back to 2017 are much appreciated. This work was supported by the Fonds de la Recherche Scientifique - FNRS
under Grants No. F.4503.20 ("HighSpinSymm") and T.0022.19 ("Fundamental issues
in extended gravitational theories").
\newpage

\appendix
\section{Conventions}\label{app:convention}
In the body of the paper, we work with four-dimensional flat space with Euclidean signature equipped with a metric 
\begin{align}
    ds^2=dx_0^2+dx_1^2+dx_2^2+dx_3^2=2dx^+dx^-+2dzd\bar{z}\,,
\end{align}
where
\begin{align}
    x^{\pm}=\frac{x^3\pm ix^0}{\sqrt{2}}\qquad \text{and}\qquad z=\frac{x^2-ix^1 }{\sqrt{2}}\,,\quad \bar{z}=\frac{x^2+ix^1}{\sqrt{2}}\,.
\end{align}
Using Pauli matrices, any vector $x^{\mu}=(x^0,x^1,x^2,x^3)$ can be written as a 2 by 2 matrix 
\begin{align}
    x^{\alpha\alpha'}= \frac{1}{\sqrt{2}}\begin{pmatrix} x^0+ix^3 & -ix^1+x^2\\
    -ix^1-x^2 & x^3-ix^3
    \end{pmatrix}=\begin{pmatrix} x^+ & z\\
    -\bar{z} & \ \ x^- \end{pmatrix} \,.
\end{align}
Here, $\alpha,\beta=0,1$ and $\alpha',\beta'=0',1'$. Obviously, $ x^{\alpha\alpha'}x_{\alpha\alpha'}=2(x^+x^-+z\bar{z})=x_0^2+x_1^2+x_2^2+x_3^2\,$. One raises and lowers spinorial indices with $SL(2,\RR)$-invariant tensors $\epsilon_{\alpha\beta}$ (or $\epsilon_{\alpha'\beta'}$) where
\begin{align}
    \epsilon_{01}=1\qquad \text{and}\qquad \epsilon_{\alpha\beta}=-\epsilon_{\beta\alpha} \,.
\end{align}
Their inverses are defined by $ \epsilon_{\alpha\gamma}\epsilon^{\beta\gamma}=\delta_{\alpha}^{\ \beta}$. One can raise and lower indices as $ x^{\alpha}=x_{\beta}\epsilon^{\alpha\beta}$ and $x_{\alpha}=x^{\beta}\epsilon_{\beta\alpha}$. Moreover, we can introduce the following inner products on the space of (un)primed indices for later convenience:
\begin{align}
    \langle xy\rangle = x^{\alpha}y_{\alpha}=x^{\alpha}y^{\beta}\epsilon_{\beta\alpha}\,,\qquad [xy]=x^{\alpha'}y_{\alpha'}=x^{\alpha'}y^{\beta'}\epsilon_{\beta'\alpha'}\,.
\end{align}
\section{Crash course on Twistor Geometry}\label{app:twistor}
\paragraph{More on twistor space.} When twistor geometry was introduced in 1967 by Penrose \cite{Penrose:1967wn}, it provided an alternative way to work with physical processes in an auxiliary space called the twistor space. Consider a 3-dimensional complex projective space, denoted $\projectivespace^3$, with homogeneous coordinates
\begin{align}
    Z^A=(Z^1,Z^2,Z^3,Z^4)\,,\quad Z^i\neq 0\,,\quad t\,Z^A\sim Z^A\quad \text{with}\quad \forall t\in \CC^*\,.
\end{align}
The projective twistor space $\PT$ is defined to be an open subset of $\projectivespace^3$ with coordinates \begin{align*}
    Z^A=(\lambda_{\alpha},\mu^{\alpha'})\,.
\end{align*}
It is well-known that $\projectivespace^3$ can be covered by two patches $\Ucal_{\pm}$ \cite{Wolf:2010av}:
\begin{align}
    \Ucal_+&=\left\{\left(\lambda_{\alpha}^+,\mu^{\alpha'}_+=\frac{\mu^{\alpha'}}{\lambda_0}\right)\Bigg|\quad \lambda_0\neq 0\,,\,\lambda_{\alpha}^+\in U_+\right\}\,,\\
    \Ucal_-&=\left\{\left(\lambda_{\alpha}^-,\mu^{\alpha'}_-=\frac{\mu^{\alpha'}}{\lambda_1}\right)\Bigg|\quad \lambda_1\neq 0\,,\,\lambda_{\alpha}^-\in U_-\right\}\,,
\end{align}
where $\lambda_{\alpha}=(\lambda_0,\lambda_1)$ are homogeneous coordinates of $\projectivespace^1$. Bases on the \textit{hat} operator in \eqref{eq:hatoperator}, the canonical choice for the coordinates on the fibres are
\begin{align}
    \lambda_{\alpha}^+=\binom{\lambda_+}{-1}\qquad \text{and}\qquad \lambda_{\alpha}^-=\binom{1}{-\lambda_-}\,.
\end{align}
Here, $\lambda^{\pm}$ are coordinates of $U_{\pm}\subset\projectivespace^1=U_+\cup U_-$. In these patches the incident relations \eqref{eq:incident} become
\begin{align}
    \mu^{\alpha'}_{\pm}=x^{\alpha\alpha'}\lambda_{\alpha}^{\pm}\,.
\end{align}
By introducing a dual twistor $\hat{Z}^A=(\hat{\lambda}_{\alpha},\hat{\mu}^{\alpha'})$ in the same patches, namely $\hat{Z}^A_{\pm}=(\hat{\lambda}_{\alpha}^{\pm},\hat{\mu}^{\alpha'}_{\pm})$, one can show that
\begin{align}
    x^{\alpha\alpha'}=\frac{\mu^{\alpha'}_{\pm}\hat{\lambda}^{\alpha}_{\pm}-\hat{\mu}^{\alpha'}_{\pm}\lambda^{\alpha}_{\pm}}{\langle \hat{\lambda}_{\pm}\lambda_{\pm}\rangle}\,,
\end{align}
where
\begin{align}
    \hat{\lambda}_+^{\alpha}=\binom{1}{\bar{\lambda}_+}\qquad \text{and}\qquad \hat{\lambda}_-^{\alpha}=\binom{\bar{\lambda}_-}{1}
\end{align}
The top-form $K$ in these patches becomes
\begin{align}
    K=\frac{d\lambda_{\pm}d\bar{\lambda}_{\pm}}{(1+\lambda_{\pm}\bar{\lambda}_{\pm})^2} \,.
\end{align}
To demonstrate how the integral over the fibers works, let us consider the patch $U_+\subset \projectivespace^1$. One can make the following change of variables $\lambda_+=re^{i\theta}$ and $\bar{\lambda}_+=re^{-i\theta}$ where $r\in \RR^+$. Then,
\begin{align}
    K=-2i\frac{rdrd\theta}{(1+r^2)^2}
\end{align}
which is in accordance with the fact that $\projectivespace^1$ is diffeomorphic to $S^2$. An important observation to prove \eqref{eq:bridge} is that the angle integral over $\theta$ cancels all components that are $\theta$-dependent. Hence, \eqref{eq:bridge} implies the following integral
\begin{align}
   -4\pi i \int \frac{rdr}{(1+r^2)^{m+2}}S_{\alpha(m)}T^{\alpha(n)}\delta_{m,n}=- \frac{2\pi i}{m+1}S_{\alpha(m)}T^{\alpha(m)}\,.
\end{align}

\paragraph{Complex Structure.} On $\PT$, an almost complex structure $J$ is defined to be a linear map $J:T(\PT)\rightarrow T(\PT)$ on the tangent bundle $T(\PT)$ of $\PT$ where $J^2=-\text{id}$. Therefore, $T(\PT)$ can be decomposed (as in usual complex geometry) into eigenspaces of $J$ that are associated with $\pm \text{i}$ eigenvalues.
Then, a $(1,0)$-vector is defined to be a vector that has eigenvalue $+\text{i}$, while a $(0,1)$-vector is a vector that has eigenvalue $-\text{i}$. Therefore, the tangent bundle $T(\PT)$ can be decomposed into
\begin{align}
    T(\PT)=T_{1,0}(\PT)\oplus T_{0,1}(\PT)\,.
\end{align}
We can define
\begin{align}
    \pl=dZ^A\frac{\pl}{\pl Z^A}\qquad \text{and}\qquad \bar{\pl}=d\bar{Z}^A\frac{\pl}{\pl\bar{Z}^A}=d\hat{Z}^A\frac{\pl}{\pl\hat{Z}^A}\,
\end{align}
to split $T(\PT)$ into $T_{1,0}(\PT)$ and $T_{0,1}(\PT)$. Note that we take $\bar{\pl}$ as our definition of complex structure on $\PT$. A complex structure is said to be integrable if $\bar{\pl}^2=0$. 

\footnotesize
\setstretch{1.0}
\bibliographystyle{JHEP-2}
\bibliography{twistor.bib}

\end{document}